\documentclass[11pt]{article}
\usepackage{hyperref}
\usepackage{amsmath}
\usepackage{graphicx,epsfig}
\usepackage{subfigure}
\usepackage{color}
\usepackage{bm}
\usepackage{multirow}
\usepackage{verbatim}
\usepackage{amssymb}
\usepackage{amsfonts}
\usepackage{amsmath}

\usepackage{authblk}

\usepackage{enumitem}

\usepackage[displaymath]{lineno}

\usepackage{multirow}

\usepackage{graphics}
\usepackage{epsfig}
\usepackage{pgf}
\usepackage{pgfpages}
\usepackage{xcolor}
\usepackage{pifont}

\graphicspath{{./Figs/}}

\usepackage{float}     % locating the picture exactly at that place

\usepackage{titletoc}
\usepackage[raggedright,pagestyles]{titlesec}

\usepackage{fancyhdr}
\usepackage{lastpage}
\usepackage{layout}
\usepackage[footnotesep=10mm]{geometry}
\usepackage{caption2}

\usepackage{setspace}

\makeatletter
\renewcommand\@biblabel[1]{\footnotesize$[{#1}]$\selectfont}
\makeatother  %

\titlecontents{part}[0em]{\vskip3mm \bf B.\contentslabel{9em}}{}
{\bf \hspace*{0em}}
{\titlerule*[0.8pc]{.}\contentspage}[\addvspace{0.5ex}]

\titlecontents{section}[1em]{\it \contentslabel{0em}}{\hspace*{1.5em}}
{\it\hspace*{2.3em}}
{\titlerule*[0.8pc]{.}\contentspage}[\addvspace{0.5ex}]

\titlecontents{subsection}[2.7em]{\contentslabel{0em}}{\hspace*{2.5em}}{\hspace*{2.3em}}
{\titlerule*[0.8pc]{.}\contentspage}[\addvspace{0.5ex}]

\titleformat{\part}{\raggedright\large\bfseries}{\bf\normalsize{B.\thepart} }{1em}{}
\titleformat{\section}{\raggedright\normalsize\bfseries}{\bf\normalsize{\S\,\thesection}.}{1em}{}
\titleformat{\subsection}{\raggedright\normalsize\bfseries}{\thesubsection}{1.5em}{}
\titleformat{\subsubsection}{\raggedright\normalsize\itshape}{\thesubsubsection}{1em}{}

\usepackage{hyperref}

\begin{document}
%\linenumbers
\title{Relativistic effect of entanglement in fermion-fermion scattering}
\author[1]{\small Jinbo Fan\thanks{fanjinbo@emails.bjut.edu.cn}}
\author[1]{\small Xinfei Li \thanks{xfli09@emails.bjut.edu.cn}}
\affil[1]{\small Institute of Theoretical Physics, Beijing University of Technology,  Beijing 100124, China}
\date{\today}
\maketitle
\begin{abstract}

\setlength{\parindent}{0pt} \setlength{\parskip}{1.5ex plus 0.5ex
minus 0.2ex} %\noindent
We study the properties of entanglement entropy among scattering particles as observed from different inertial moving frames, based on an exemplary QED process $e^+e^-\rightarrow\mu^+\mu^-$. By the explicit calculation of the Wigner rotation, the entanglement entropy of scattering particles is found to be Lorentz invariant. We also study the behavior of the entanglement between spin degrees of freedom for scattering particles in moving frames. 
This quantity, being found to change with different inertial reference frames, does not exhibit as a Lorentz invariant.
\end{abstract}

%%%%%%%%%%%%%%%%%%%%%%%%%%%%%%%%%%%%%%%%%%%%%%%%%%%%%%%
%%%%%%%%%%%%%%%%%%%%%%%%%%%%%%%%%%%%%%%%%%%%%%%%%%%%%%%
%%%%%%%%%%%%%%%%%%%%%%%%%%%%%%%%%%%%%%%%%%%%%%%%%%%%%%%

\vspace*{10mm}

\section{Introduction}

Quantum information which originally formulated in terms of nonrelativistic quantum mechanics, recently has been receiving many interests within the more fundamental framework of quantum field theory \cite{Cardy:2004,Calabrese:2009,Horodecki:2009}, since studying quantum information theoretic issues provides some insights into the structure of interacting quantum field theories, in particular enriching the understanding of conformal field theories via calculating the entanglement among different regions. To analyze the structure of interacting quantum field theory, Refs. \cite{Balasubramanian:2011wt,Hsu:2012gk} studied the entanglement entropy of two divided momentum spaces with the perturbative calculations method, and this method was then followed by Refs. \cite{Park:2014hya} for the study of the entanglement between two scalar particles in the scattering process in a weakly coupled field theory.

The physics of scattering process plays a crucial role in a wide variety of physical experiments, which probe the behavior of elementary particles. In a scattering process, the final state is determined by an initial state and S-matrix. Thus, the final state are entangled states of in-fields and one are interested in quantifying the entanglement entropy generated by the elastic collision of two initial particles. The relation between the scattering and the entanglement entropy in scalar field theory are also studied in Refs. \cite{ Lello:2013bva,Peschanski:2016hgk,Carney:2016tcs,Grignani:2016igg}, in terms of different S-matrix formalisms which play a significant role in these computation.

Lately, Ref. \cite{Fan:2017hcd} extends the study to the relativistic regime involving fermions. The entanglement of a two-fermion state is generally not Lorentz invariant although certain states and partitions are. This occurs due to the spin degree of freedom of a particle mixing with momentum under Lorentz transformation \cite{Gingrich:2002ota}. Peres et al. \cite{Peres:2002} have argued that the spin entropy determined by the reduced density matrix tracing out the momentum is not invariant under Wigner rotation. Czachor et al. \cite{Czachor:2003,Czachor:2005} have demonstrated that the difficulty for define spin reduced density matrices in a Lorentz invariant manner arises from the dependence of the Wigner rotation on the momentum of the state. Some researches \cite{Gingrich:2002ota,Friis:2009va,He:2007gk} concentrating on the entanglement of biparticle system obtained the results, the entanglement between total degrees of freedom for two particles, i.e., including momentum and spin, is Lorentz invariant, and their spin or momentum entanglement change from when one reference frame to another.

In this paper, using the perturbative method in Refs. \cite{Balasubramanian:2011wt,Park:2014hya}, we also study the entanglement for spin degrees of freedom of particles in the fermion-fermion scattering process with a weak coupling.
It is convenient to study an elastic scattering in two-particle Fock space, where incoming and outgoing particles can be regarded as asymptotically free particles. In order to measures the spin entanglement between two particles, we study mutual information in momentum space, $I(s_A,s_B)=S(s_A)+S(s_B)-S(s_{A\cup B})$, where $S(s_A)$ is the von Neumann entropy of the reduced density matrix of spin degrees of freedom for particle $A$, and $S(s_{A\cup B})$ is the von Neumann entropy of the reduced density matrix of total spin degrees of freedom for two particle.

The structure of this work is organized as follows. In section 2 the Wigner rotation is briefly reviewed, and the properties of entanglement of scattering particles in moving frame are studied. In section 3 how the spin entanglement varies in the relativistic frame is carefully analyzed. At last, conclusion is made in section 4.

\section{Entanglement entropy of scattering particles}

Refs. \cite{Park:2014hya, Fan:2017hcd} studied the entanglement entropy between two particles in an elastic scattering process with a particular perturbative technique. It was found that the change in entanglement entropy from incoming particles to outgoing particles is proportional to cross section in a weak coupling theory. In order to ascertain the relation between Lorentz symmetry and the entanglement of scattering particles in a weak coupling, we calculate this quantity in different inertial reference frames.
In our context, we adopt Weinberg's notation \cite{Weinberg}:  metric signature, index ordering and gamma matrix.

\subsection{ Reduced density matrix}
Before starting to study the entanglement of scattering particles in moving frame, let us recall what Refs. \cite{Fan:2017hcd} studied. At weak coupling, we can assume the unitarity of local interaction terms to be guaranteed at lower orders of perturbation \cite{Park:2014hya}. The initial and final states  can be viewed as superposition of the basis of free Hamiltonian $H_{\textup{free}}$ so that we can divide the total Hilbert space as $\mathcal{H}_{\textup{tot}}=\mathcal{H}_A\otimes\mathcal{H}_B$. For an elastic scattering process of two fermions in 2-particle Fock space,  incoming and outgoing particle states  can be described as
\begin{align}
\vert p,s;q,r\rangle=\sqrt{2E_{\textbf{p}}}~{a^s_{\textbf{p}}}^{\dagger}\vert0\rangle_A\otimes \sqrt{2E_{\textbf{q}}}~{b^r_{\textbf{q}}}^{\dagger}\vert0\rangle_B,
\end{align}
where $\textbf{p}$ and $\textbf{q}$ are the 3-momenta of particles, and $s$, $r$ denote the spin of particles.
The fermionic creation/annihilation operators obey the commutation relations,
\begin{align}
\{a^s_{\textbf{p}}, {a^r_{\textbf{k}}}^\dagger\}=(2\pi)^3\delta^{(3)}(\textbf{p}-\textbf{k})\delta^{sr}
,~~~~
\{b^n_{\textbf{q}}, {b^m_{\textbf{\textit{l}}}}^\dagger\}=(2\pi)^3\delta^{(3)}(\textbf{q}-\textbf{\textit{l}})\delta^{nm}.
\end{align}
Then the inner product between 2-particle states is defined as
\begin{align}
\langle k,s^\prime; l,r^\prime\vert p,s;q,r\rangle=2E_{\textbf{k}}2E_{\textbf{\textit{l}}}(2\pi)^3\delta^{(3)}(\textbf{k}-\textbf{p})  (2\pi)^3\delta^{(3)}(\textbf{\textit{l}}-\textbf{q})\delta^{ss^\prime}\delta^{rr^\prime}.
\end{align}

One chooses a general initial state,
\begin{align}
\label{ini}
\vert\textup{ini}\rangle=\sum_{\sigma_1,\sigma_2}a_{\sigma_1\sigma_2}\vert p,\sigma_1;q,\sigma_2\rangle,
\end{align}
where the coefficients satisfy $\sum_{\sigma_1,\sigma_2}a_{\sigma_1\sigma_2}a^{\dagger}_{\sigma_1\sigma_2}=1$. The final state is determined by the initial state and the $S$ matrix \cite{Park:2014hya},
\begin{align}
\label{fin}
\vert \textup{fin}\rangle=\int\frac{d^3\textbf{k}}{(2\pi)^3}\frac{1}{2E_{\textbf{k}}}\frac{d^3\textbf{\textit{l}}}{(2\pi)^3}\frac{1}{2E_{\textbf{\textit{l}}}}
\sum_{\sigma_3,\sigma_4}\sum_{\sigma_1,\sigma_2}a_{\sigma_1\sigma_2}
\vert k,\sigma_3;l,\sigma_4\rangle\langle k,\sigma_3;l,\sigma_4\vert\textbf{S}\vert p,\sigma_1;q,\sigma_2\rangle.
\end{align}
The $T$ matrix is defined as
\begin{align}
i\textbf{T}=\textbf{S}-\textbf{1},~~~~\langle k,\sigma_3;l,\sigma_4\vert i\textbf{T}\vert p,\sigma_1;q,\sigma_2\rangle=(2\pi)^4\delta^{(4)}(p+q-k-l)\times i\mathcal{M}_{\sigma_1\sigma_2\sigma_3\sigma_4}(p,q\rightarrow k,l),
\end{align}
where $\mathcal{M}_{\sigma_1\sigma_2\sigma_3\sigma_4}$ is the invariant matrix element in a scattering process.

Introduce a shorthand notation
\begin{align}
\mathcal{M}_{\sigma_3\sigma_4}(a)\equiv
\sum_{\sigma_1\sigma_2} a_{\sigma_1\sigma_2}\mathcal{M}_{\sigma_1\sigma_2\sigma_3\sigma_4}(p,q\rightarrow k,l),
\end{align}
thus the final state is simply written as
\begin{align}
\vert \textup{fin}\rangle=&\sum_{\sigma_1,\sigma_2}a_{\sigma_1\sigma_2}\vert p,\sigma_1;q,\sigma_2\rangle
\\ \notag
&+i \sum_{\sigma_3,\sigma_4}\int\frac{d^3\textbf{k}}{(2\pi)^32E_{\textbf{k}}}\frac{d^3\textbf{\textit{l}}}{(2\pi)^32E_{\textbf{\textit{l}}}}
(2\pi)^4\delta^{(4)}(p+q-k-l)
\mathcal{M}_{\sigma_3\sigma_4}(a)
\vert k,\sigma_3;l,\sigma_4\rangle.
\end{align}
The setting of the entire scattering process is designated to occur in a large spacetime volume of duration $T$ and spatial volume $V$; these factors are artifact caused by regulating delta functions Ref.\cite{Weinberg},
\begin{align}
\label{deltafnc}
\delta^3_V(\textbf{p}-\textbf{p}^{\prime})=\frac{V}{(2\pi)^3}\delta_{\textbf{p},\textbf{p}^{\prime}},~~~~\delta_T(E_{\textup{if}})\equiv\delta_T(E_{\textup{fin}}-E_{\textup{ini}})=\frac{1}{2\pi}\int_{-T/2}^{T/2}dt~e^{i(E_{\textup{fin}}-E_{\textup{ini}})t},
\end{align}
which implies $V=(2\pi)^3\delta_V^{(3)}(0)$ and $T=(2\pi)\delta_T(0)$, respectively.

From Eq. $(\ref{fin})$ we can evaluate the total density matrix of final state by $\rho_{AB}=\vert \textup{fin}\rangle\langle\textup{fin}\vert$. The reduced density matrix
$\rho^{(\textup{fin})}_A$ is obtained by tracing out the degrees of freedom for particle $B$,
$\rho^{(\textup{fin})}_A=\mathcal{N}^{-1}~\textup{tr}_B~\rho_{AB}$,  producing the following result
\begin{align}
\rho^{\textup{(fin)}}_A=&\frac{1}{\mathcal{N}}\biggr\{\sum_{\sigma_1,\sigma^\prime_1}\mathcal{I}_{\sigma_1\sigma^\prime_1}2E_{\textbf{q}}V
\vert p,\sigma_1\rangle\langle p,\sigma^\prime_1\vert
\\ \notag
+\lambda^2&\sum_{\sigma_3,\sigma^{\prime}_3}
\int_{\textbf{k}\neq\textbf{p}}
\frac{\{2\pi\delta(E_{\textup{if}})\}^2}{2E_{\textbf{k}}2E_{\textbf{p}+\textbf{q}-\textbf{k}}2E_{\textbf{k}}}
\mathcal{A}_{\sigma_3\sigma^\prime_3}
\vert k,\sigma_3\rangle\langle k,\sigma^{\prime}_3\vert\biggr\},
\end{align}
where $\mathcal{N}$ is the normalization factor fixed by $tr_A\rho^{\textup{(fin)}}_A=1$,
\begin{align}
\mathcal{N}&=2E_{\textbf{q}}2E_{\textbf{p}}V^2+\lambda^2\int\limits_{\textbf{k}\neq\textbf{p}}
\frac{\{2\pi\delta(E_{\textup{if}})\}^2V}{2E_{\textbf{k}}2E_{\textbf{p}+\textbf{q}-\textbf{k}}}
\mathcal{A}_{\sigma_3\sigma_3},
\end{align}
and using two shorthand notations for the long expressions
\begin{align}
\mathcal{I}_{\sigma_1\sigma^\prime_1}=\sum_{\sigma_2}a_{\sigma_1\sigma_2}a^{\star}_{\sigma^\prime_1\sigma_2}
,~~~~
\mathcal{A}_{\sigma_3\sigma^\prime_3}=\frac{1}{\lambda^2}\sum_{\sigma_4}\mathcal{M}_{\sigma_3\sigma_4}(a)
\mathcal{M}^{\star}_{\sigma^\prime_3\sigma_4}(a).
\end{align}

In the weak coupling, the reduced density matrix for particle $A$ at order $\lambda^2$ can be written as
\begin{align}
\label{rhoA}
\rho^{\textup{(fin)}}_A=\textup{diag}\left((1-\lambda^2 \mathcal{A})\mathcal{I},...,\lambda^2 \mathcal{A}_k,...\right),
\end{align}
where the elements of this matrix correspond to $\frac{\vert p,\sigma_1\rangle\langle p,\sigma^\prime_1\vert}{2 E_{\textup{p}}V}, \cdots,\frac{\vert k,\sigma_1\rangle\langle k,\sigma^\prime_1\vert}{2 E_{\textup{k}}V},\cdots$, and
\[\mathcal{I}=\begin{pmatrix}
\mathcal{I}_{11}  & \mathcal{I}_{12}  \\
\mathcal{I}_{21}  & \mathcal{I}_{22} \\
\end{pmatrix}
,~~~~
\mathcal{A}_k=\frac{\{2\pi\delta(E_{\textup{if}})\}^2}{2E_{\textbf{k}}2E_{\textbf{q}}2E_{\textbf{p}}2E_{\textbf{p}+\textbf{q}-\textbf{k}}V^2}
\begin{pmatrix}
\mathcal{A}_{11}     &  \mathcal{A}_{12} \\
\mathcal{A}_{21}     &  \mathcal{A}_{22} \\
\end{pmatrix},\]
\begin{align}
\mathcal{A}=\int\limits_{\textbf{k}\neq\textbf{p}}
\frac{\{2\pi\delta(E_{\textup{if}})\}^2}{2E_{\textbf{k}}2E_{\textbf{q}}2E_{\textbf{p}}2E_{\textbf{p}+\textbf{q}-\textbf{k}}V}\mathcal{A}_{\sigma_3\sigma_3}.
\end{align}
Then the entanglement entropy between particle $A$ and $B$ in the final is $S^{\textup{(fin)}}_E=-tr\rho^{\textup{(fin)}}_A\log\rho^{\textup{(fin)}}_A$.

\subsection{Wigner Rotations}

Before considering the properties of the entanglement entropy between two fermions in moving frame, it is necessary to take a brief review on the effect of a single fermion state under Lorentz transformation. As shown in Ref. \cite{Alsing:2002}, for a massive particle, $p^2<0$, we can choose a standard four-momentum $k^\mu$ (usually taken in the particle's rest frame, $k^\mu=(0,0,0,1)$), and express any $p^\mu$ of this class as
\begin{align}
p^\mu=L^\mu_{~\nu}(p)k^{\nu},
\end{align}
where $L^{\mu}_{~\nu}(p)$ are some standard Lorentz transformation that depend on $p^\mu$,  taking the four-momentum $k\rightarrow p$. In terms of standard momentum states $\vert k,\sigma\rangle$, the corresponding state-vector can be defined as
\begin{align}
\vert p,\sigma\rangle\equiv U(L(p))\vert k,\sigma\rangle,
\end{align}
where
\begin{align}
P^\mu\vert k,\sigma\rangle&=k^\mu\vert k,\sigma\rangle,
\\ \notag
J^2\vert k,\sigma\rangle&=s(s+1)\vert k,\sigma\rangle,
\\ \notag
J_z\vert k,\sigma\rangle&=\sigma\vert k,\sigma\rangle,
\end{align}
where $s$ and $\sigma$ are the spin and the $z$ component of the spin for the particle, respectively.

The effect on the state-vector $\vert p,\sigma\rangle$ under an arbitrary Lorentz transformation $\Lambda$ (rotation and boost) is
\begin{align}
U(\Lambda)\vert p,\sigma\rangle=U(L(\Lambda p))U(W(\Lambda,p))\vert k,\sigma\rangle,
\end{align}
where $W(\Lambda,p)\equiv L^{-1}(\Lambda p)\Lambda L(p)$. The transformation takes $k$ to $L(p)k=p$, and then to $\Lambda p$, and then back to $k$, so it must be a rotation. These rotations are called the Wigner rotations, which act only on the spin component $\sigma$ in rest frame. Hence, the state becomes,
\begin{align}
\label{Lorentz}
U(\Lambda)\vert p,\sigma\rangle=\sum_{\sigma^\prime}D^{(s)}_{\sigma^\prime\sigma}(W(\Lambda,p))\vert k,\sigma^\prime\rangle,
\end{align}
where $D^{(s)}_{\sigma^\prime\sigma}(W(\Lambda,p))$ is the spin $s$ representation of the rotation $W(\Lambda,p)$. The Wigner rotation matrices are given by Ref. \cite{Alsing:2002, Edmonds}
\[
D^{(j_n)}_{\sigma^\prime\sigma}(W(\Lambda,p))=\begin{pmatrix}
\cos(\Omega_{\textbf{p}}/2) & -\sin(\Omega_{\textbf{p}}/2)  \\
\sin(\Omega_{\textbf{p}}/2)  &  \cos(\Omega_{\textbf{p}}/2)
\end{pmatrix},
\]
where the  Wigner rotation angle $\Omega_{\textbf{p}}$ depends on the direction and the rapidity of the boost.

Thus, under Lorentz transformation $\Lambda$, the transformed spinor $u^{\Lambda}(p,\sigma)$ can be  rewritten as a Wigner rotation of the spinors $u(p_{\Lambda},\sigma)$,
\begin{align}
u^{\Lambda}(p,1/2)&\equiv \cos(\Omega_{\textbf{p}}/2) u(p_{\Lambda},1/2)+\sin(\Omega_{\textbf{p}}/2) u(p_{\Lambda},-1/2),
\\ \notag
u^{\Lambda}(p,-1/2)&\equiv -\sin(\Omega_{\textbf{p}}/2)u(p_{\Lambda},1/2)+ \cos(\Omega_{\textbf{p}}/2)u(p_{\Lambda},-1/2).
\end{align}

\subsection{Entanglement entropy of scattering particles in moving frame}

For an observer in moving frame, the final state under Lorentz transformation $\Lambda$ is
\begin{align}
\vert\textup{fin}\rangle^{\Lambda}=\sum_{\sigma_1,\sigma_2}a_{\sigma_1\sigma_2}\vert p,\sigma_1;q,\sigma_2\rangle^{\Lambda}+i \sum_{\sigma_3\sigma_4}\int d\Pi_2
\mathcal{M}_{\sigma_3\sigma_4}^{\Lambda}(a)
\vert k,\sigma_3;l,\sigma_4\rangle^{\Lambda},
\end{align}
which has the Lorentz invariance structure
\begin{align}
\int d\Pi_2=\int\frac{d^3\textbf{k}}{(2\pi)^3}\frac{1}{2E_{\textbf{k}}}\frac{d^3\textbf{\textit{l}}}{(2\pi)^3}\frac{1}{2E_{\textbf{\textit{l}}}}(2\pi)^4\delta^{(4)}(p+q-k-l).
\end{align}
The general state $\vert k,\sigma_3;l,\sigma_4\rangle^{\Lambda}$ appears to be transformed by Wigner rotation,
\begin{align}
\label{finlam}
\vert k,\sigma_3;l,\sigma_4\rangle^{\Lambda}=\sum_{\sigma^\prime_3,\sigma^\prime_4}D_{\sigma^\prime_3\sigma_3}(W(\Lambda,k))D_{\sigma^\prime_4\sigma_4}(W(\Lambda,l))\vert k_\Lambda,\sigma^\prime_3;l_\Lambda,\sigma^\prime_4\rangle.
\end{align}
In the folloing we denote $D_{\sigma^\prime\sigma}(p)\equiv D_{\sigma^\prime\sigma}(W(\Lambda,p))$. Then the final state is reinterpreted as
\begin{align}
\vert\textup{fin}\rangle^{\Lambda}=&\sum_{\sigma^\prime_1,\sigma^\prime_2}\sum_{\sigma_1,\sigma_2}a_{\sigma_1\sigma_2}
D_{\sigma^\prime_1\sigma_1}(p)D_{\sigma^\prime_2\sigma_2}(q)
\vert p_{\Lambda},\sigma^\prime_1;q_{\Lambda},\sigma^\prime_2\rangle
\\ \notag
&+i\sum_{\sigma_3^{\prime},\sigma_4^{\prime}}\sum_{\sigma_3,\sigma_4}
\int d\Pi_2~
D_{\sigma_3^{\prime}\sigma_3}(k)D_{\sigma_4^{\prime}\sigma_4}(l)
\mathcal{M}^{\Lambda}_{\sigma_3\sigma_4}(a)
\vert k_\Lambda,\sigma_3^{\prime}; l_\Lambda,\sigma_4^{\prime}\rangle,
\end{align}
It is worth to note that
\begin{align}
\notag
\mathcal{M}_{\sigma_1\sigma_2\sigma_3\sigma_4}^{\Lambda}(k,l\rightarrow p,q)=\sum_{\sigma^\prime_3,\sigma^\prime_4}\sum_{\sigma^\prime_1,\sigma^\prime_2}
D^{-1}_{\sigma_3\sigma^\prime_3}(k)D^{-1}_{\sigma_4\sigma_4^\prime}(l)
D_{\sigma_1^\prime\sigma_1}(p)D_{\sigma_2^\prime\sigma_2}(q)
\mathcal{M}_{\sigma^\prime_1\sigma^\prime_2\sigma^\prime_3\sigma^\prime_4}(k_\Lambda , l_\Lambda \rightarrow p_\Lambda ,q_\Lambda).
\end{align}

For simplicity, we define
\begin{align}
\label{SimbD}
b_{\sigma_1^\prime\sigma_2^\prime}=&\sum_{\sigma_1,\sigma_2}a_{\sigma_1\sigma_2}D_{\sigma_1^\prime\sigma_1}(p)D_{\sigma_2^\prime\sigma_2}(q),
\notag \\
\mathcal{M}^{\Lambda}_{\sigma^\prime_3\sigma^\prime_4}(b)=&
\sum_{\sigma_3^{\prime\prime},\sigma_4^{\prime\prime}}
D_{\sigma_3^{\prime}\sigma_3}(k)D_{\sigma_4^{\prime}\sigma_4}(l)
\mathcal{M}^{\Lambda}_{\sigma_3\sigma_4}(a).
\end{align}
Thus the final state can be expressed as
\begin{align}
\vert\textup{fin}\rangle^{\Lambda}=&\sum_{\sigma_1,\sigma_2}b_{\sigma_1\sigma_2}\vert p_\Lambda,\sigma_1;q_\Lambda,\sigma_2\rangle
\notag \\
+&i\sum_{\sigma_3,\sigma_4}\int\frac{d^3\textbf{k}}{(2\pi)^3}\frac{2\pi\delta(E_{\textup{if}})}{2E_{\textbf{k}}2E_{\textbf{p}+\textbf{q}-\textbf{k}}}\mathcal{M}^{\Lambda}_{\sigma_3\sigma_4}(b)
\vert k_\Lambda,\sigma_3;p_\Lambda+q_\Lambda-k_\Lambda,\sigma_4\rangle.
\end{align}

 In the analogous calculation, the reduced density matrix for one particle under the action of Lorentz transformation $\Lambda$ has the following results:
\begin{align}
\rho^{(\Lambda,\textup{fin})}_A&=\sum_{\sigma_1,\sigma^\prime_1}\sum_{\sigma_2}b_{\sigma_1\sigma_2}b^{\dagger}_{
\sigma^\prime_1\sigma_2}2E_{\textbf{q}_\Lambda}V_{q_\Lambda}
\vert p_\Lambda,\sigma_1\rangle\langle p_\Lambda,\sigma^\prime_1\vert
\\ \notag
+&\sum_{\sigma_3,\sigma^\prime_3}\int_{\textbf{k}\neq\textbf{p}}\frac{d^3\textbf{k}}{(2\pi)^3}\frac{(2\pi)\delta(E_{\textup{if}})}{2E_{\textbf{k}}2E_{\textbf{p}+\textbf{q}-\textbf{k}}}
\frac{(2\pi)\delta(E^\Lambda_{\textup{if}})}{2E_{\textbf{k}_\Lambda}}
\left\{\sum_{\sigma_4}\mathcal{M}^{\Lambda}_{\sigma_3\sigma_4}(b)(\mathcal{M}^{\Lambda}_{\sigma^\prime_3\sigma_4}(b))^{\dagger}\right\}
\vert k_\Lambda,\sigma_3\rangle\langle k_\Lambda,\sigma^\prime_3\vert,
\end{align}
Note that regularization factors $T_{\textbf{p}}$ and $V_{\textbf{p}}$ are not Lorentz invariant. Since these quantities  $E_{\textbf{p}}\delta^{(3)}(\textbf{p}-\textbf{q})$ and $(2\pi)^{(4)}(p+q-k-l)$ are Lorentz invariant, it is easy to obtain
\begin{align}
E_{\textbf{p}}V_{\textbf{p}}=E_{\textbf{p}_\Lambda}V_{\textup{p}_\Lambda},~~~~T_{\textbf{p}}V_{\textbf{p}}=T_{\textbf{p}_\Lambda}V_{\textbf{p}_\Lambda}.
\end{align}
Thus,  the Lorentz-transformated reduced density matrix $\rho^{(\Lambda,\textup{fin})}$ can be derived as
\begin{align}
\label{rhoALam}
\rho^{(\Lambda,\textup{fin})}_A&=\frac{1}{\mathcal{N}^{\Lambda}}\biggr\{
\sum_{\sigma_1,\sigma^\prime_1}\mathcal{I}^{\Lambda}_{\sigma_1\sigma^\prime_1}
2E_{\textbf{q}}V_{\textbf{q}}
\vert p_\Lambda ,\sigma_1\rangle\langle p_\Lambda,\sigma^\prime_1\vert
\\ \notag
&+\lambda^2\sum_{\sigma_3,\sigma^\prime_3}\int_{\textbf{k}\neq\textbf{p}}\frac{d^3\textbf{k}}{(2\pi)^3}\frac{\{2\pi\delta(E_{\textup{if}})\}^2}{2E_{\textbf{k}}2E_{\textbf{p}+\textbf{q}-\textbf{k}}2E_{\textbf{k}}}
\mathcal{A}^{\Lambda}_{\sigma_3\sigma^\prime_3}
\vert k_\Lambda,\sigma_3\rangle\langle k_\Lambda,\sigma^\prime_3\vert
\biggr\},
\end{align}
where
\begin{align}
\mathcal{I}^{\Lambda}_{\sigma_1\sigma^\prime_1}=\sum_{\sigma_2}b_{\sigma_1\sigma_2}b^{\dagger}_{
\sigma^\prime_1\sigma_2},~~~~
\mathcal{A}^{\Lambda}_{\sigma_3\sigma^\prime_3}=\frac{1}{\lambda^2}\sum_{\sigma_4}\mathcal{M}_{\sigma_3\sigma_4}^{\Lambda}(b)(\mathcal{M}^{\Lambda}_{\sigma^\prime_3\sigma_4}(b))^{\dagger}.
\end{align}
and $\mathcal{N}^{\Lambda}$ is the normalization factor fixed by $tr_A\rho^{(\Lambda,\textup{fin})}_A=1$.

In the weak coupling, the Lorentz-transformed reduced density matrix at order $\lambda^2$ can be written as
\begin{align}
\label{rhoAlam}
\rho^{(\Lambda,\textup{fin})}_A=\textup{diag}\left((1-\lambda^2 \mathcal{A}^{\Lambda})\mathcal{I}^{\Lambda},...,\lambda^2 \mathcal{A}^{\Lambda}_k,...\right),
\end{align}
where the elements of this matrix correspond to $\frac{\vert p_\Lambda,\sigma_1\rangle\langle p_\Lambda,\sigma^\prime_1\vert}{2 E_{\textup{p}}V}, \cdots,\frac{\vert k_\Lambda,\sigma_1\rangle\langle k_\Lambda,\sigma^\prime_1\vert}{2 E_{\textup{k}}V},\cdots$, and
\[\mathcal{I}^{\Lambda}=\begin{pmatrix}
\mathcal{I}^{\Lambda}_{11}  &  \mathcal{I}^{\Lambda}_{12}  \\
\mathcal{I}^{\Lambda}_{21}  &  \mathcal{I}^{\Lambda}_{22} \\
\end{pmatrix},~~~~
\mathcal{A}^{\Lambda}_k
=\frac{\{2\pi\delta(E_{\textup{if}})\}^2}{2E_{\textbf{q}}2E_{\textbf{p}}2E_{\textbf{k}}2E_{\textbf{p}+\textbf{q}-\textbf{k}}V^2}
\begin{pmatrix}
\mathcal{A}^{\Lambda}_{11}      &  \mathcal{A}^{\Lambda}_{12} \\
\mathcal{A}^{\Lambda}_{21}      &  \mathcal{A}^{\Lambda}_{22} \\
\end{pmatrix},\]
\begin{align}
\mathcal{A}^{\Lambda}&=\int\limits_{\textbf{k}\neq\textbf{p}}
\frac{d^3\textbf{k}}{(2\pi)^3}
\frac{\{2\pi\delta(E_{\textup{if}})\}^2}{2E_{\textbf{q}}2E_{\textbf{p}}2E_{\textbf{k}}2E_{\textbf{p}+\textbf{q}-\textbf{k}}V}
\mathcal{A}^{\Lambda}_{\sigma_3\sigma_3}.
\end{align}

Following the spirit of Ref.\cite{Fan:2017hcd},  the scattering $e^+e^-\rightarrow\mu^+\mu^-$ will be taken into consideration  for simplicity. Since only $s-$channel contributes to this process, we choose the following initial state with parametrization of the entanglement between the spin degrees of freedom
\begin{align}
\label{inistate}
\vert\textup{ini}\rangle=\cos\eta\vert p,1/2; q,1/2\rangle+\sin\eta~ e^{i\beta}\vert p,-1/2; q,-1/2\rangle,
\end{align}
with $\sigma=\pm1/2$ representing the spin quantized along the $z$-axis, $\eta\in[0,\pi/2]$ parametrizing  the spin entanglement of the state, and $\beta\in[-\pi/2,3\pi/2]$ labelling the relative phase of the superposed states $\vert p,1/2; q,1/2\rangle$ and $\vert p,-1/2; q,-1/2\rangle$.

To obtain a more explicit formula, we could make the simplest choice of evaluating entanglement entropy in the center of mass frame S. The mass of electronic is ignorable, since the ratio $m_e/m_{\mu}\simeq1/200$ is much smaller than the fractional error introduced by neglecting higher-order terms in the perturbation series. The initial and final four momenta for $e^+e^-\rightarrow\mu^+\mu^-$ are
\begin{align}
p^\mu&=(0,0,E,E),~~~~q^\mu=(0,0,-E,E),
\\ \notag
k^\mu&=(\sqrt{E^2-m^2}\sin\theta,0,\sqrt{E^2-m^2}\cos\theta,E),
\\ \notag
l^\mu&=(-\sqrt{E^2-m^2}\sin\theta,0,-\sqrt{E^2-m^2}\cos\theta,E).
\end{align}

It is interesting to study the properties of the entanglement entropy of scattering particles in moving frames, because spin and momentum become mixed when viewed by a moving observer \cite{Gingrich:2002ota,Peres:2004}. Since any transformation matrix $D_{\sigma^\prime\sigma}(p)$ does not depend on $\textbf{p}$ for a pure rotation \cite{Gingrich:2002ota}, it is sufficient to focus on pure boosts. Without loss of generality, we make a boost to a frame $S^\prime$ travelling in the $z$ direction with rapidity $w$,
\[
\Lambda=\begin{pmatrix}
1 & 0 & 0 & 0 \\
0 & 1 & 0 & 0  \\
0 & 0 & \cosh w & \sinh w  \\
0 & 0 & \sinh w & \cosh w  \\
\end{pmatrix},
\]
where $w\rightarrow0$ means that the reference frame $S^\prime\rightarrow S$.

For the chosen initial state, the matrices in Eqs.\eqref{rhoA} and \eqref{rhoAlam}
 \[
\mathcal{I}=
\begin{pmatrix}
\sin^2\eta & 0 \\
0 & \cos^2\eta \\
\end{pmatrix}
,~~~~
\mathcal{I}^{\Lambda}=\begin{pmatrix}
\frac{1}{2}(1+\cos2\eta\cos2\Omega_{\textbf{p}}) & \frac{1}{2}\cos2\eta\sin\Omega_{\textbf{p}} \\
\frac{1}{2}\cos2\eta\sin\Omega_{\textbf{p}} & \frac{1}{2}(1-\cos2\eta\cos2\Omega_{\textbf{p}}) \\
\end{pmatrix},
\]
 have the same eigenvalues, $\sin^2\eta$ and $\cos^2\eta$. The corresponding entanglement entropy of incoming particles in different inertial reference frame are
 \begin{align}
S^{(\textup{ini})}_E=S^{(\Lambda,\textup{ini})}_E=-\sin^2\eta\log\sin^2\eta-\cos^2\eta\log\cos^2\eta.
\end{align}

At the lowest order, the scattering amplitude in the reaction $e^+e^-\rightarrow\mu^+\mu^-$ can be written as
 \begin{align}
\mathcal{M}_{\sigma_1\sigma_2\sigma_3\sigma_4}(pq\rightarrow kl)=-\frac{\lambda}{4E}\bar{v}(q,\sigma_2)\gamma^\mu u(p,\sigma_1)\bar{u}(k,\sigma_3)\gamma_\mu v(l,\sigma_4),
\end{align}
where mode functions $u(p,\sigma)$ and $v(p,\sigma)$ are chosen as momentum eigenstates in Ref.$\cite{Alsing:2002}$.
Using the given four-momentum of scattering particles and Lorentz transformation $\Lambda$, it is easy to compute Wigner angle $\Omega_{\textbf{p}}$ and  Wigner rotation matrices $D_{\sigma^\prime\sigma}(p)$  by Ref. \cite{Alsing:2002}.  Even though scattering amplitude $\mathcal{M}_{\sigma_1\sigma_2\sigma_3\sigma_4}$ is different from $\mathcal{M}^{\Lambda}$ viewed by the Lorentz-boosted observer, we find
\begin{align}
\mathcal{A}=\mathcal{A}^{\Lambda}=\frac{T}{V}\frac{\sqrt{E^2-m^2}(2E^2+m^2+(E^2-m^2)\cos\beta\sin2\eta)}{24\pi E^5},
\end{align}
where the factors $T$ and $V$ are constants which denote the value of $T_{\textbf{p}}$ and $V_{\textbf{p}}$ in center of mass frame.
Furthermore, the matrices $\mathcal{A}_k$ and $\mathcal{A}^{\Lambda}_k$ in Eqs.$(\ref{rhoA},\ref{rhoAlam})$ have the same roots, the corresponding entangelment entropy of outgoing particles viewed by different observers have the same result, $S^{(\textup{fin})}_E=S^{(\Lambda,\textup{fin})}_E$.

Thus, we  find that entanglement entropy of scattering particles for initial and final state are Lorentz invariant. This is consistent with the fact that a local unitary transformation will not affect any measure of entanglement \cite{Vidal:1998re, Parker:1999kd}.

The change in the entanglement entropy of scattering particles in an arbitrary inertial reference frame is
\begin{align}
\triangle S_E
&=-\lambda^2\log\lambda^2\mathcal{A}-(1-\lambda^2\mathcal{A})(\cos^2\eta\log\cos^2\eta+\sin^2\eta\log\sin^2\eta)+\lambda^2\mathcal{A}
\\ \notag
&-\frac{\lambda^2\sqrt{E^2-m^2}T}{128\pi^2 E^3V}\frac{}{ }\int d\Omega(a_{k1}\log(a_{k1})+a_{k2}\log(a_{k2}))++\lambda^2\mathcal{A}\log(\frac{16E^4V^2}{T^2}).
\end{align}
At the lowest, $\lambda^2$, the cross section for the scattering $e^+e^-\rightarrow\mu^+\mu^-$ is a Lorentz invariant,
\begin{align}
\sigma_{\vert\textup{ini}\rangle}=\frac{\lambda^2\sqrt{E^2-m^2}(2E^2+m^2+(E^2-m^2)\cos\beta\sin2\eta)}{48\pi E^5}.
\end{align}
From the values of $\mathcal{A}$ and $\sigma_{\vert\textup{ini}\rangle}$, we can further confirm the conclusion in Ref. \cite{Park:2014hya, Fan:2017hcd} that the change in entanglement entropy of scattering particles  in a weak coupling theory is proportional to cross section.

\begin{figure}[H]
  \centering
  \includegraphics[scale=0.8]{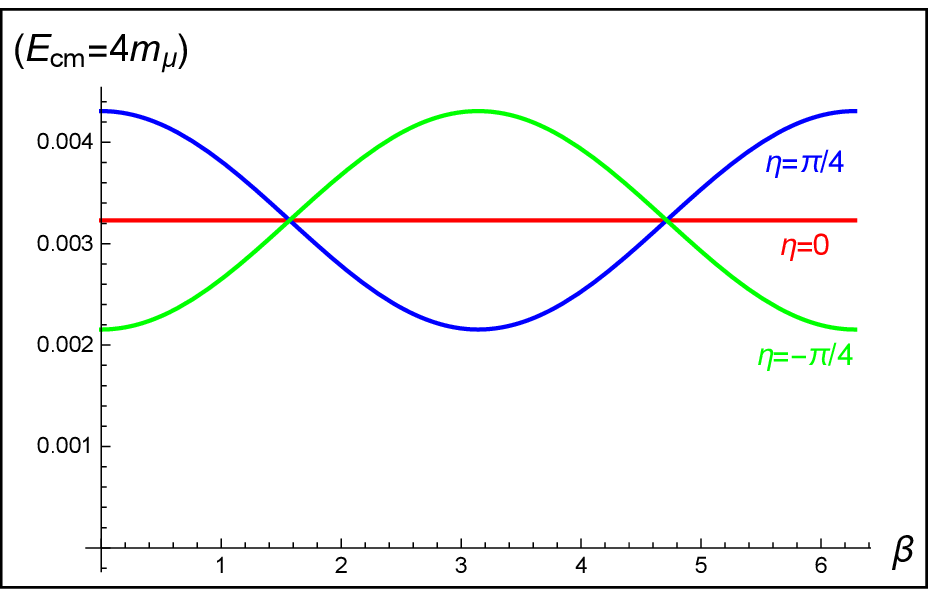}
 \caption{\textup{The $\lambda^2\log\lambda^2$ order contribution to the change in entanglement entropy of scattering fermions as a function of entangle parameter} $\eta$\textup{ and}\textup{ phase factor} $\beta$. $E_{\textup{cm}}$ \textup{is the total initial energy}.}
\end{figure}

Finally, the leading perturbative contribution of the change in entanglement entropy of scattering particles is
\begin{align}
\label{ent}
\triangle S_E=-\lambda^2\log\lambda^2\mathcal{A}+O(\lambda^2),
\end{align}
where the leading contribution to spin entanglement entropy is $\lambda^2\log\lambda^2$ and vanishes as $\lambda\rightarrow0$. This is consistent with the fact that interactions cause change in the degree of entanglement between subsystem \cite{Balasubramanian:2011wt, Kumar:2017ctm}. From \eqref{ent}, $\triangle S_E>0$, interaction between two particles could  increase the degree of entanglement of scattering particles, which is analogous to increase of entropy principle in thermodynamics.

The change in entanglement entropy of scattering fermions in the process $e^+e^-\rightarrow\mu^+\mu^-$ is plotted in Fig.1.
The figure describes a function about the dependence of the change in entanglement entropy on phase factor $\beta$ for initial states of different degrees of entanglement, $\eta=0$ (not entangled), $\eta=\pi/4$ (symmetric Bell state) and $\eta=-\pi/4 $(antisymmetric Bell state). The  figure tells that, for a initial state of certain degree of entanglement, in addition to the effect of interaction, the interference of different histories corresponding to the superposition state also change the entanglement entropy of scattering particles.

\section{Spin entanglement of scattering particles in moving frame}

In the previous calculation, the initial state and the final state of two fermions can be regarded as generated by the basis of an asymptotically free Hamiltonian, then their total Hilbert space can be divided into $\mathcal{H}_{tot}=\mathcal{H}_A\otimes\mathcal{H}_B$. For a subsystem, say $A$, choosing spin state as the complete basis, its Hilbert space $\mathcal{H}_A$ could be further divided into $\mathcal{H}_{p_{A}}\otimes\mathcal{H}_{s_A}$, $\textit{i.e.}$, the spin and momentum degrees of freedom of subsystem $A$.
After performing the same decomposition for subsystem $B$, the total Hilbert space for the initial state and final state can be decomposed into $\mathcal{H}_{p_A}\otimes\mathcal{H}_{s_A}\otimes\mathcal{H}_{p_B}\otimes\mathcal{H}_{s_B}$. It is interesting to investigate a particular quantity, the mutual information between spin degrees of freedom for two fermionic field,
\begin{align}
\label{mutual}
I(s_A,s_B)=S(s_A)+S(s_B)-S(s_A\cup s_B),
\end{align}
which  measures the spin entanglement between two particles. Here $S(X)$ is the Von Neumann entropy of the reduced density matrix of subsystem $X$. Furthermore, the properties of the mutual information in moving frame will be studied.

 In the following, we calculate the spin entanglement entropy of outgoing particles in the scattering process $e^+e^-\rightarrow\mu^+\mu^-$ ,with the same initial spin state parametrization as $(\ref{inistate})$,
 \begin{align}
\vert\textup{ini}\rangle=\cos\eta\vert p,1/2; q,1/2\rangle+\sin\eta~ e^{i\beta}\vert p,-1/2; q,-1/2\rangle.
\end{align}

As explained in the previous section, the final state is determined by the initial state and $S$ matrix. From Eq.$(\ref{rhoALam})$, the reduced density matrix $\rho^{(\Lambda,\textup{fin})}_{s_A}$ is obtained by tracing out the momentum degrees of freedom for particle $A$, $\rho^{(\Lambda,\textup{fin})}_{s_A}=tr_{p_A}\rho^{(\Lambda,\textup{fin})}_A$, producing the following results at order $\lambda^2$
\begin{align}
\label{rhosAlam}
\rho^{(\Lambda,\textup{fin})}_{s_A}=(1-\lambda^2 \mathcal{A}^{\Lambda})I^{\Lambda}+\lambda^2\mathcal{B}^{\Lambda},
\end{align}
where
\[I^{\Lambda}=\begin{pmatrix}
I^{\Lambda}_{11}  &  I^{\Lambda}_{12}  \\
I^{\Lambda}_{21}  &  I^{\Lambda}_{22} \\
\end{pmatrix}
,~~~~
\mathcal{B}^\Lambda=\int\limits_{\textbf{k}\neq\textbf{p}}
\frac{d^3\textbf{k}}{(2\pi)^3}
\frac{\{2\pi\delta(E_{\textup{if}})\}^2}{2E_{\textbf{q}}2E_{\textbf{p}}2E_{\textbf{k}}2E_{\textbf{p}+\textbf{q}-\textbf{k}}V}
\begin{pmatrix}
\mathcal{A}^{\Lambda}_{11}      &  \mathcal{A}^{\Lambda}_{12} \\
\mathcal{A}^{\Lambda}_{21}      &  \mathcal{A}^{\Lambda}_{22} \\
\end{pmatrix}, \]
\begin{align}
\label{AK}
\mathcal{A}^{\Lambda}=\int\limits_{\textbf{k}\neq\textbf{p}}
\frac{d^3\textbf{k}}{(2\pi)^3}
\frac{\{2\pi\delta(E_{\textup{if}})\}^2}{2E_{\textbf{q}}2E_{\textbf{p}}2E_{\textbf{k}}2E_{\textbf{p}+\textbf{q}-\textbf{k}}V}
\mathcal{A}^{\Lambda}_0.
\end{align}

Both entanglement parameter $\eta$ and phase factor $\beta$ can control the degree of entanglement for initial state, while we choose $\beta$ as zero for simplicity. In the analogous calculation, the eigenvalues of the spin reduced density matrix $\rho^{(\Lambda,\textup{fin})}_{s_A}$ can readily be obtain,
\begin{align}
\notag
r^{\Lambda}_{k1}&=\sin^2\eta+\frac{T\lambda^2\sqrt{E^2-m^2}}{192\pi VE^5}
((8\cos2\eta-8+3b^{\Lambda}_{22}-16\cos\eta\sin^3\eta)E^2-2(-1+\cos2\eta+\sin2\eta)^2m^2),
\\ \notag
r^{\Lambda}_{k2}&=\cos^2\eta-\frac{T\lambda^2\sqrt{E^2-m^2}}{192\pi VE^5}((8-3b^{\Lambda}_{11}+8\cos2\eta+16\cos\eta^3\sin\eta)E^2+2(1+\cos\eta-\sin2\eta)^2m^2),
\end{align}
where $b^{\Lambda}_{ij}=\int d\theta\mathcal{A}^{\Lambda}_{ij}$. The corresponding entanglement entropy is
\begin{equation}
S^{(\Lambda,\textup{fin})}(s_A)=\begin{cases}
-\cos^2\eta\log(\cos^2\eta)-\sin^2\eta\log(\cos^2\eta)+\frac{\lambda^2T}{V}f_1(\eta,\omega)+O(\lambda^4), & \textup{if}~~\eta\neq0; \\
 -\frac{\lambda^2\log\lambda^2T}{V}f_2(\eta,\omega)+O(\lambda^2), & \textup{if}~~ \eta=0.
 \end{cases}
 \end{equation}
where $f_1$ and $f_2$ are functions of parameter $\eta$ and $\omega$ for the given particle energy $E$ and mass $m$. Obviously, $S^{(\Lambda,\textup{fin})}(s_B)$ has the same value as $S^{(\Lambda,\textup{fin})}(s_A)$.

The total spin reduced density matrix for two particles at $\lambda^2$ order can be written as
\begin{align}
\rho^{(\Lambda,\textup{fin})}_s=(1-\lambda^2 \mathcal{A})\mathcal{C}^{\Lambda}+\lambda^2\mathcal{D}^\Lambda,
\end{align}
where
\[\mathcal{C}^{\Lambda}=\begin{pmatrix}
\mathcal{I}^{\Lambda}_{1111}  & \mathcal{I}^{\Lambda}_{1112}  &  \mathcal{I}^{\Lambda}_{1121}  & \mathcal{I}^{\Lambda}_{1122} \\
\mathcal{I}^{\Lambda}_{1211}  & \mathcal{I}^{\Lambda}_{1212}  &  \mathcal{I}^{\Lambda}_{1221}  & \mathcal{I}^{\Lambda}_{1222} \\
\mathcal{I}^{\Lambda}_{2111} & \mathcal{I}^{\Lambda}_{2112}  &  \mathcal{I}^{\Lambda}_{2121}  & \mathcal{I}^{\Lambda}_{2122} \\
\mathcal{I}^{\Lambda}_{2211} & \mathcal{I}^{\Lambda}_{2212}  &  \mathcal{I}^{\Lambda}_{2221}  & \mathcal{I}^{\Lambda}_{2222} \\
\end{pmatrix},\]
\begin{align}
\mathcal{A}^{\Lambda}=\int\limits_{\textbf{k}\neq\textbf{p}}
\frac{d^3\textbf{k}}{(2\pi)^3}
\frac{\{2\pi\delta(E_{\textup{if}})\}^2}{2E_{\textbf{k}}2E_{\textbf{p}}2E_{\textbf{q}}2E_{\textbf{p}+\textbf{q}-\textbf{k}}V}
\mathcal{A}^{\Lambda}_0,
\end{align}
\[\mathcal{D}^\Lambda=\int\limits_{\textbf{k}\neq\textbf{p}}
\frac{d^3\textbf{k}}{(2\pi)^3}
\frac{\{2\pi\delta(E_{\textup{if}})\}^2}{2E_{\textbf{k}}2E_{\textbf{q}}2E_{\textbf{p}}2E_{\textbf{p}+\textbf{q}-\textbf{k}}V}
\begin{pmatrix}
\mathcal{A}^{\Lambda}_{1111}  & \mathcal{A}^{\Lambda}_{1112}  &  \mathcal{A}^{\Lambda}_{1121}  & \mathcal{A}^{\Lambda}_{1122} \\
\mathcal{A}^{\Lambda}_{1211}  & \mathcal{A}^{\Lambda}_{1212}  &  \mathcal{A}^{\Lambda}_{1221}  & \mathcal{A}^{\Lambda}_{1222} \\
\mathcal{A}^{\Lambda}_{2111} & \mathcal{A}^{\Lambda}_{2112}  &  \mathcal{A}^{\Lambda}_{2121}  & \mathcal{A}^{\Lambda}_{2122} \\
\mathcal{A}^{\Lambda}_{2211} & \mathcal{A}^{\Lambda}_{2212}  &  \mathcal{A}^{\Lambda}_{2221}  & \mathcal{A}^{\Lambda}_{2222} \\
\end{pmatrix}, \]
and
\begin{align}
\mathcal{I}^{\Lambda}_{\sigma_1\sigma_2\sigma^\prime_1\sigma^\prime_2}&=b_{\sigma_1\sigma_2}b^{\dagger}_{\sigma^\prime_1\sigma^\prime_2}
,~~~~
\mathcal{A}^{\Lambda}_{\sigma_3\sigma^\prime_3\sigma_4\sigma^\prime_4}=
\frac{1}{\lambda^2}\mathcal{M}_{\sigma_3\sigma_4}^{\Lambda}(\mathcal{M}^{\Lambda}_{\sigma^\prime_3\sigma^\prime_4})^{\dagger}.
\end{align}
Similarly, the entanglement entropy between spin and momentum for outgoing particles  is derived as
\begin{align}
S^{(\Lambda,\textup{fin})}(s)=-\frac{\lambda^2\log\lambda^2T}{V}f_4(\eta,\omega)+O(\lambda^2).
\end{align}

Finally, according to Eq.$(\ref{mutual})$, the spin entanglement entropy of outgoing particles is
\begin{equation}
\label{mutlam}
I^{(\Lambda,\textup{fin})}=\begin{cases}
-\frac{\lambda^2\log\lambda^2T}{V}f_3(\eta,\omega)+O(\lambda^2), & \textup{if}~~\eta=0; \\
-2\cos^2\eta\log(\cos^2\eta)-2\sin^2\eta\log(\cos^2\eta)+\frac{\lambda^2\log\lambda^2T}{V}f_4(\eta,\omega)+O(\lambda^2), & \textup{if}~~\eta\neq0.
\end{cases}
\end{equation}
where
\begin{align}	
\notag
f_3&=\frac{\sqrt{E^2-m^2}}{128\pi E^3}(2b^{\Lambda}_{22}-(d^{\Lambda}_{1212}+d^{\Lambda}_{2121}+d^{\Lambda}_{2222})),
\\ \notag
f_4&=\frac{\sqrt{E^2-m^2}}{128\pi E^3}(d^{\Lambda}_{1111}+2(d^{\Lambda}_{1212}+d^{\Lambda}_{2121})+d^{\Lambda}_{2222}+(-d^{\Lambda}_{1111}+d^{\Lambda}_{2222})\cos2\eta-(d^{\Lambda}_{1122}+d^{\Lambda}_{2211})\sin2\eta),
\end{align}
with $d^{\Lambda}_{ijkl}=\int d\theta \mathcal{A}^{\Lambda}_{ijkl}$. The functions $f_3$ and $f_4$ appearing in the entanglement entropy for $e^+e^-\rightarrow\mu^+\mu^-$ is plotted in Fig.2.

\begin{figure}[htp]
  \centering
  \vspace{0pt}
  \includegraphics[scale=0.8]{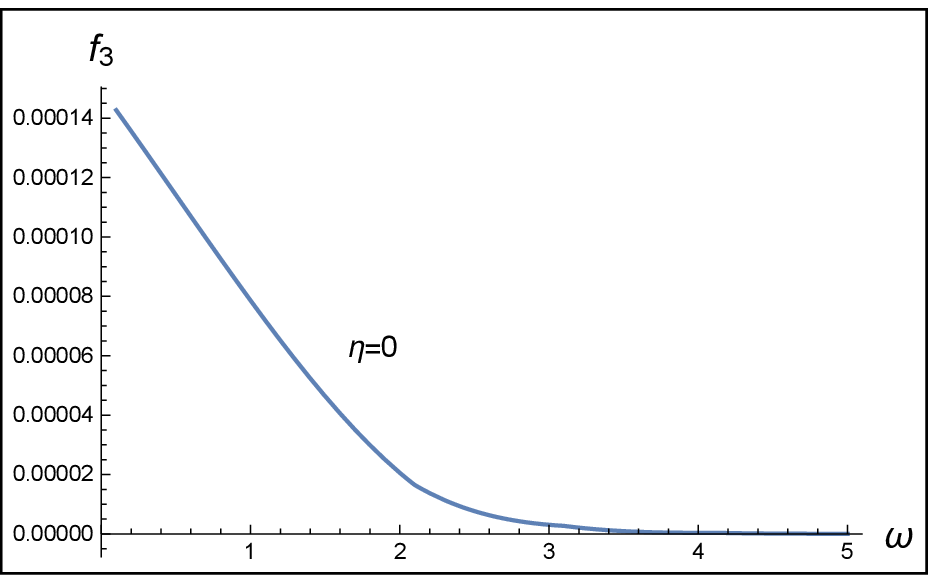}
  \qquad
  \qquad
  \vspace{0pt}
  \includegraphics[scale=0.8]{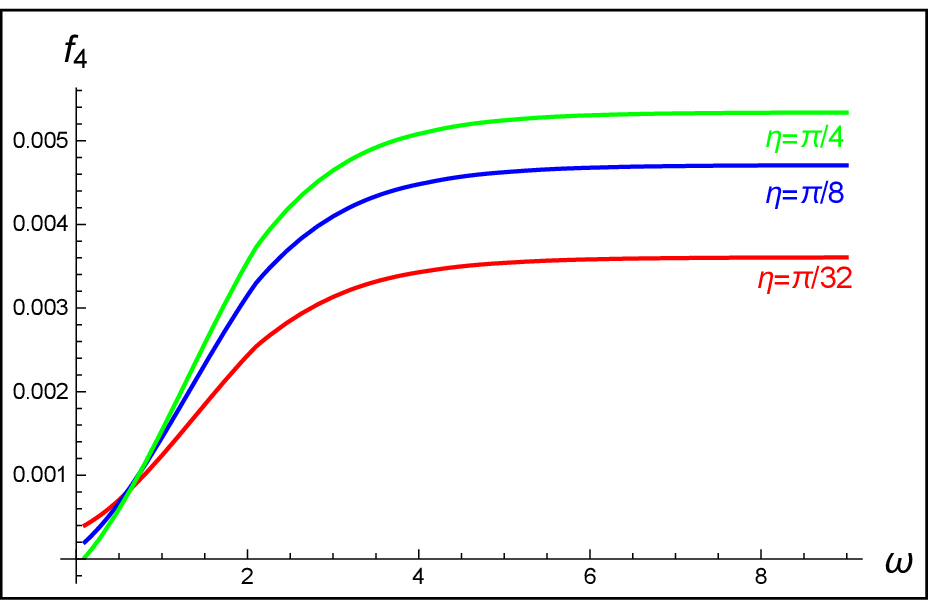}
 \caption{\textup{ The $\lambda^2\log\lambda^2$ order contribution to spin entanglement entropy of outgoing fermions as a function of rapidity $\omega$. The left figure is for the initial state with parameter $\eta=0$ (not entangled). The right figure is for the initial state with $\eta=\pi/32, \pi/8, \pi/4$}. }
\end{figure}

According to Eq. \eqref{mutlam} and Fig. 2, the spin entanglement of outgoing particles  in the moving frame  monotonously decreases with the increasing of rapidity $\omega$, which indicates that the spin entanglement entropy is not Lorentz invariant. This variation shows that the Lorentz boost may induce the decoherence of the spin entanglement. While this does not mean that a Lorentz transformation always decreases the spin entanglement entropy of outgoing particles. One approach to obtain the reverse effect is to apply the appropriate boost along other axis.
 As mentioned in Ref.\cite{Gingrich:2002ota,Peres:2002}, as the spin and momentum degrees of freedom being entangled by Lorentz boosts,  an increasing in spin entanglement would occur at the expense of momentum entanglement, since the entanglement of scattering particles is Lorentz invariant.
In the limit $\omega\rightarrow \infty$ (boost to the speed of light), the spin entanglement entropy of outgoing particles reaches a constant that only depends on the degree of entanglement for incoming particles. In particular, if the initial state is not entangled, the spin entanglement entropy of the final state will vanish when viewed from a Lorentz-transformed frame along the $z$ direction.

\section{Conclusion}

We study the properties of entanglement entropy among scattering particles as observed from different inertial moving frames, based on an exemplary QED process $e^+e^-\rightarrow\mu^+\mu^-$. By explicit calculation of the Wigner rotation, two inertial frames moving with constant relative velocity report the same result for the the entanglement entropy of scattering particles. This result further confirms the conclusion in Ref.\cite{Park:2014hya,Fan:2017hcd} that the change in entanglement entropy in a weak coupling theory is proportional to cross section which is a Lorentz invariant. Combining the supports from studies in Refs.\cite{Peschanski:2016hgk,Carney:2016tcs,Grignani:2016igg} on the relation between entanglement entropy of scattering particles and cross section in other contexts, our work indicates that the entanglement between total degrees of freedom for two particles, including momentum and spin, is Lorentz invariant. We also study the mutual information between spin degrees of freedom for two fermionic field substitutes for spin entanglement entropy of scattering particles. This quantity, being calculated in moving frame and found to change with different inertial reference frames, does not exhibit as a Lorentz invariant. 

Although the investigations are based on the scattering process $e^+e^-\rightarrow\mu^+\mu^-$, similar results can be expected for general fermion-fermion scattering with a weak coupling. It would be interesting to generalize our computations to higher orders in perturbation theory which considers the runing of coupling constant. Corresponding investigations in a strong coupling field theory is also expected to advance the further study on the relation between the scattering and the entanglement entropy. In the AdS/CFT correspondence, both of the scattering amplitude and entanglement entropy in a strongly coupled field theory are associated with minimal surfaces in a bulk gravity theory \cite{Janik:2000aj,Ryu:2006bv}. It is also interesting to attempt for the holographic understanding  of the relation between the scattering and entanglement entropy by means of AdS/CFT.

\section{Acknowledgment}
The authors would like to thank Xiang-Dong Gao, Yanbin Deng, Wenyu Wang, Ding-fang Zeng and Yong-Chang Huang for useful discussions. This work is supported by the National Natural Science Foundation of China (Grants No. 11275017 and No. 11173028).

\end{document}